Original Article

# A reliability data aggregation method for fusion-oriented research facilities

Helena Ortiz [a], Juan Eloy Ruiz-Castro [a], Christian Acal [a,*], Karol Kowal [b,c], Juan Chiachío [d], Ana María Aguilera [a]

[a] Department of Statistics and O.R. and IMAG, University of Granada, Spain
[b] National Centre for Nuclear Research (NCBJ), 7 Andrzeja Sotana Str., PL-05-400 Otwock, Poland
[c] Faculty of Physics, University of Warsaw (UW), 5 Ludwika Pasteura Str., PL-02-093 Warsaw, Poland
[d] Department Structural Mechanics and Hydraulics Engineering and DaSCI, University of Granada, Spain

ABSTRACT

This paper aims to develop a statistical method for integrating multiple incomplete data sources to improve the estimation of the failure-rate probability distribution of fusion-specific devices. The data aggregation approach proposed in this work is to support the Reliability, Availability, Maintainability, and Inspectability (RAMI) studies for fusion-oriented research facilities. This work focuses on developing a procedure for handling missing or incomplete data from multiple sources, as well as formulating a statistical framework for combining data from different sources that refer to the same equipment type and failure mode. Specifically, a Bayesian result is adopted under the assumption of a constant failure rate over time. This framework motivates the use of the Gamma distributions to model the failure-rate parameter of each device. Subsequently, two alternative procedures are introduced to approximate, using a single Gamma distribution, the weighted mixture formed by combining Gamma distributions corresponding to the same equipment type across multiple data sources. The proposed method is validated through an extensive simulation study and further illustrated by application to the accelerator systems of the IFMIF-DONES facility. Finally, the proposed algorithm has been implemented in the open-source software R to facilitate its use in the nuclear fusion community.

## 1. Introduction

Ensuring Reliability, Availability, Maintainability, and Inspectability (RAMI) has been one of the major challenges facing fusion technology for the last few decades [1]. Initial studies focusing on the operational experience related to the fusion device availability included the tokamaks JET (UK), AUG (Germany), Tore Supra (France), DIII-D (USA), and TFTR (USA), as well as the tritium facilities TLK (Germany), TSTA (USA), and TPL (Japan) [2]. The frameworks for RAMI studies have been established at the early design phase of new facilities oriented toward nuclear fusion, ITER [3], DEMO [4], HL-2M [5], and EAST [6]. Alongside the development of RAMI simulation methods, the equipment reliability data has been collected from both fission and fusion facilities. However, while data from commercial power plants was systematically recorded and rigorously reported, research-oriented fusion facilities generally applied less stringent practices for failure monitoring and statistical data management. Although significant efforts have been made to establish a standardized Fusion Component Failure Rate Database (FCFRDB) [7], the available multi-source equipment reliability datasets remain highly diverse and heterogeneous, leading to increased uncertainty in RAMI models for emerging fusion research facilities. This work aims to present a general method for aggregating multi-source reliability data for the fusion-specific components to support RAMI analyses of fusion-oriented research facilities. As a reference case, the reliability data for selected equipment of the IFMIF-DONES accelerator systems is considered.

Numerous efforts have been undertaken within the EUROfusion Early Neutron Source Work Package (WPENS) [8–10] to develop RAMI models for the International Fusion Materials Irradiation Facility-Demo Oriented Neutron Energy Source (IFMIF-DONES) [11,12]. These models are central to demonstrating the feasibility of the 70% operational availability target set for the facility, which entails two planned shutdown periods per year where deferred corrective maintenance activities need to be accommodated with predetermined and statutory maintenance activities. The reliability of such assessments is therefore critically dependent on the quality and consistency of the failure rate data used as inputs. However, these RAMI studies usually progress at different levels of maturity and are carried out by multiple expert teams from both industry and academia, each focusing on specific subsystems. To derive meaningful and consistent results, it is essential to integrate the models developed for the individual systems. In this sense, one of the principal challenges associated with such integration is the harmonization of the input datasets.

* Corresponding author.
*E-mail address:* chracal@ugr.es (C. Acal).

In RAMI studies, component failure rates are commonly estimated using heterogeneous datasets extracted from multiple sources that differ in content, structure, and level of accuracy. This situation gives rise to two main issues: (1) when large datasets are available, more than one value may be identified as a reference for defining the failure rates of typical components; and (2) no single database is comprehensive, as certain component types may be absent or the available information may be limited, for example, to mean values only. In general, two approaches can be adopted to address this issue: (1) evaluating the relevance of the available data sources or records and selecting a single reference dataset; and (2) combining data from multiple sources or records, thereby integrating several inputs into a unified failure-rate model. The second approach appears more justified, as it enables the consideration of data diversity reflecting the wide range of operating conditions to which the equipment may be subjected. However, this approach requires a well-defined procedure for the selection and integration of the most relevant inputs. When properly implemented, such a procedure can prevent situations in which different analysts select different input data for the same components across various systems. This concern is particularly relevant in IFMIF-DONES, where the logistics and maintenance analysis is conducted simultaneously with the design of the building and plant systems by multiple teams from industry and academia, requiring periodic coordination to ensure consistency across interfaces [13].

Recently, in the context of fusion research, a novel method was established to integrate the heterogeneous reliability data for fusion-specific components of the DONES Accelerator Systems [14]. In this approach, the parameters of the failure rate density function associated with each relevant dataset are estimated. Then, a Monte Carlo simulation is performed to randomly sample the failure rate for a given component across multi source inputs. An empirical probability distribution is finally constructed from the simulated failure rates, assuming equal contribution from all data sources.

The implementation of the multi-source data integration algorithm was motivated by the need to update the RAMI studies for the accelerator systems initiated in the earlier stages of the project [15–19]. Then, the accelerator design was iteratively refined to maximize operational availability while incorporating the most relevant operational data. As a result, the design evolved into an engineering solution ready for construction [20]. However, despite these efforts, the accelerator remains the most challenging part of the IFMIF-DONES installation for achieving the specified availability target [12]. To select a strategy for optimizing reliability allocation, component-level reliability models must be refined.

In this line, the present research aims to introduce a novel statistical methodology with a solid theoretical basis that, firstly, avoids the need to impose additional restrictive assumptions when reconstructing failure-rate probability distributions from different databases in which the available information is often partial or unspecified, and, secondly, follows the recommendation of simulating different numbers of values per study and assigning weights according to the relative importance of each study by employing distribution fusion methodologies instead of Monte Carlo methods [21]. It is well-known that unweighted Monte Carlo simulation analyses, where the same number of random samples are generated from different distributions and subsequently combined into a single dataset (simulation and pooling), may yield a mixture distribution with artificially heavy tails, which may not accurately reflect reality or the relative confidence associated with each study.

Then, the well-known Poisson-Gamma framework is adopted to identify and estimate the failure rate probability distribution for each relevant data source from missing or incomplete data. Regarding the second goal, in reliability analysis, several methodological frameworks have been proposed for the integration of multiple probability density functions (PDFs), reflecting different sources of uncertainty, data, or expert knowledge (see, e.g., [22]). These approaches can be broadly classified into model blending techniques, Bayesian updating strategies, nonparametric data-driven methods, and functional data analysis (FDA) approaches. Model blending encompasses mixture models, where candidate PDFs are combined through weighted arithmetic or geometric schemes to capture heterogeneous failure mechanisms, as well as model averaging approaches that aggregate reliability measures using weights derived from information criteria or Bayesian evidence to reduce model misspecification risk [21,23]. Bayesian updating methods treat the candidate PDFs as stochastic realizations informed by prior distributions and observational data, yielding posterior distributions that balance prior knowledge and empirical evidence, particularly in data-scarce settings [24]. Nonparametric approaches, including kernel density estimation, empirical distribution fusion, and information-theoretic methods based on divergence measures such as the Kullback–Leibler (KL) divergence, offer flexible integration strategies that avoid restrictive parametric assumptions while minimizing information loss [25,26]. Finally, FDA-based methods interpret PDFs as smooth curves (functions) within a functional space, enabling the use of metrics such as the Wasserstein distance to define representative densities via Fréchet means, or to rank and combine densities using functional depth measures that provide robustness against outliers and model misspecification [27–29].

From an applied standpoint, no single method dominates. The choice depends on the data context (quantity, heterogeneity, type of uncertainty) and the application objective (safety margins, best-fit modeling, robustness). Among the methodologies reviewed, two have been selected as the most suitable for integrating the initial probability density functions.

The remainder of the manuscript is organized as follows. Section 2 presents the proposed methodology. Section 3 describes a simulation study conducted to assess the performance of the methodology. Section 4 provides an application based on a real dataset. A broader discussion and potential future directions are presented in Section 5. Finally, the main conclusions are summarized in Section 6.

## 2. Methodology

This section presents the reliability data aggregation method from multiple and incomplete data sources to estimate the failure rate probability distribution of fusion-specific equipment. This methodology consists of three well-differentiated phases: (1) selection of relevant records from multi-source databases, (2) estimation of the failure-rate probability distribution function for each of the selected records, and (3) integration of the multi-source heterogeneous data records.

### 2.1. Selection of relevant data records from the multi-source databases

The first step of the reliability data aggregation method proposed in this work involves selecting relevant records corresponding to a specific equipment type and failure mode from multi-source databases. This process is typically performed by a single expert, who explores the available databases for relevant equipment records and evaluates whether the operational context and usage conditions are comparable to the target application. This approach introduces substantial subjectivity, particularly for novel and highly specialized technical systems of fusion-oriented research facilities. In such cases, it may be difficult to justify whether the selected reference data are truly representative when only some relevant parameters are comparable. This may lead to two misleading effects: (1) overly conservative selection — when the expert chooses only a very limited number of records, thus reducing the statistical robustness of the data sampling process, and (2) overly broad selection — when the expert attempts to maximize sample size by including many records that are similar in certain aspects but substantially different in others, thus reducing the validity of the resulting model.

To mitigate such effects, the data selection process can be structured as a facilitated workshop involving a panel of experts, with clearly defined procedures for resolving disagreements and making final decisions when individual opinions differ. In such situations, determining whose judgment should be prioritized may be particularly challenging, especially when all experts possess comparable levels of expertise and scientific credibility. To address this issue, a facilitator should be appointed to guide the expert panel toward consensus by ensuring that the evaluation process remains structured, transparent, and focused on reaching agreement, without contributing technical judgments.

Another critical issue concerns the selection of parameters used to assess whether data records are sufficiently comparable to the target equipment, thus enabling the identification of a representative data sample for a specific equipment operating within specific conditions. Accordingly, the selected records should reflect technical/design similarity (such as physical dimensions, performance characteristics, operating capacity, mechanical architecture, and underlying physical and chemical processes), operational similarity (operating temperature, rotational speed, voltage, vibration levels, environmental exposure) and exploitation similarity (continuous operation, cyclic, standby or redundant). To improve consistency, this assessment may be formalized through a numerical scoring system. For example, assigning a score of 10 to records representing identical design and operating conditions and a score of 1 to those representing completely different states. Furthermore, the relative importance of individual parameters may be established to enable weighted scoring based on their significance.

The proposed procedure consists of the following steps:

1. A panel of experts is assembled and coordinated by a designated facilitator.
2. Through consensus, the panel determines which records are evaluated.
3. Each selected record is assessed using a structured similarity scoring approach.
4. Scores assigned by individual experts are weighted and averaged for each record.
5. A threshold score is determined to select the records for further analysis.

The final step requires balancing two competing objectives: increasing sample size while maintaining sample validity. A larger dataset may improve statistical representation, but only if the included records remain sufficiently relevant to the target application. This structured approach improves repeatability, reduces individual bias, and increases the transparency of failure rate data selection for further steps of the data aggregation process.

### 2.2. Failure rate distribution estimation from heterogeneous and incomplete data sources

The objective is to identify and estimate the failure rate probability distribution for each relevant data source from missing or incomplete data. For this purpose, the well-known Poisson-Gamma framework is adopted, in which the Gamma distribution provides a natural probabilistic model for the failure-rate parameter. This provides a common probabilistic framework for representing heterogeneous reliability information from different studies and facilitates the subsequent aggregation process.

The random variable of interest is the failure rate, which is assumed to be constant. Under this assumption, it is well known that a constant failure rate is always associated with an Exponential distribution for the failure time. As a consequence, the number of failures per unit of time (frequency of failures) follows a Poisson probability distribution, whose parameter is the failure rate, thereby describing the inherent randomness of the failure occurrence process. Within the Poisson-Gamma framework, the uncertainty associated with the value of the failure-rate parameter is represented by a Gamma distribution [30,31]. Therefore, the reliability information reported by each study is represented by estimating the parameters of an appropriate Gamma distribution from the available summary statistics.

Formally, let $\Lambda$ be a random variable representing the failure rate. The uncertainty associated with $\Lambda$ is modeled by a Gamma distribution, denoted as $\Lambda \sim \Gamma(\alpha,\beta); \alpha>0, \beta>0$. The probability density function is given by the following formula

$$f(\lambda;\alpha,\beta)=\frac{1}{\beta^{\alpha}\Gamma(\alpha)}\lambda^{\alpha-1}e^{-\lambda/\beta},$$

where the cumulative density function can be derived as

$$F(\lambda;\alpha,\beta)=\int_0^{\lambda} f(t;\alpha;\beta)dt.$$

It is well known that $E[\Lambda]=\mu=\alpha\beta$ and then, $\beta=\mu/\alpha$. Thus, defining $Md=\lambda_{50}$, $LB=\lambda_5$, $UB=\lambda_{95}$ and $EF=UB/Md$

$$f(\lambda;\alpha,\beta)=\frac{1}{\beta}f\left(\frac{\lambda}{\beta};\alpha,1\right)\rightarrow \lambda_p(\alpha,\beta)=\beta\lambda_p(\alpha,1),$$

with $\lambda_5$, $\lambda_{50}$ and $\lambda_{95}$ being the 5th, 50th and 95th percentiles, respectively.

Depending on the available percentiles (in RAMI studies, typically only the mean value is available, while none or only some of the aforementioned percentiles are reported), $\alpha$ is estimated from:

- None specified: the error factor is fixed as $EF=5$.
- One percentile:

$$F(\lambda_p;\alpha,\mu/\alpha)=\frac{p}{100}.$$

- Two percentiles: For example, if $LB$ and $UB$ are known

$$\frac{\lambda_{95}(\alpha,1)}{\lambda_5(\alpha,1)}=\frac{UB}{LB}.$$

- Three percentiles: The difference ratio is used; e.g.

$$\frac{\lambda_{95}(\alpha,1)-\lambda_{50}(\alpha,1)}{\lambda_{95}(\alpha,1)-\lambda_5(\alpha,1)}=\frac{UB-Md}{UB-LB}.$$

### 2.3. Integrating multi-source heterogeneous data for the same component

This objective consists of proposing a statistical methodology for integrating the probability distributions estimated for the different relevant data sources, in only one distribution that represents the variability of the failure rate for each component. The idea is to integrate the Gamma distributions estimated for the different relevant data sources, in only one Gamma distribution that represents the variability of the failure rate for each component (see [32] for a detail study of this problem).

Formally, it is assumed that each study $i=1,\ldots,n$ provides a Gamma distribution denoted as $\Lambda_i \sim \Gamma(\alpha_i,\beta_i)$, where $\alpha_i$ and $\beta_i$ are the shape and scale parameters, respectively. These specific distributions are considered independent and the objective is to derive a unique distribution $\Lambda \sim \Gamma(\alpha,\beta)$ representing the population model that generated the failure data for each individual study. Hereinafter, two different methodologies are proposed in which the population distribution $\Lambda \sim \Gamma(\alpha,\beta)$ is approximated from a weighted mixture of the individual distributions.

#### 2.3.1. Method of moments

The objective is to combine several independent Gamma distributions originated from different studies into a single Gamma distribution which is approximated from the mixture

$$\Lambda \approx \sum_{i=1}^{n} \Lambda_i \mathbf{1}_{\{Z=i\}},$$

being the discrete variable $Z$ such that $P(Z=i)=w_i$, and $\mathbf{1}_{\{Z=i\}}$ denotes the indicator function, equal to 1 when $Z=i$ and 0 otherwise.

This moment-based methodology consists of identifying for the population a Gamma distribution whose first and second order moments are the ones of the mixture.

It is well known that for a $\Lambda \sim \Gamma(\alpha, \beta)$,

$$\begin{aligned} \mu &= \mathbb{E}[\Lambda] = \alpha\beta \\ \sigma^2 &= \mathrm{Var}(\Lambda) = \alpha\beta^2 \\ m_2 &= \mathbb{E}[\Lambda^2] = \sigma^2 + \mu^2, \end{aligned}$$

with $m_2$ being the second raw (non-central) moment. For weights $w_i \geq 0$ with $\sum_{i=1}^n w_i = 1$, the mixture moments can be deduced as follows:

$$\begin{aligned} \mu &= \sum_{i=1}^n w_i \mu_i \\ \mathbb{E}[\Lambda^2] &= \int_0^\infty \lambda^2 \sum_{i=1}^n w_i f_{\Lambda_i}(\lambda; \alpha_i, \beta_i) d\lambda = \sum_{i=1}^n w_i \int_0^\infty \lambda^2 f_{\Lambda_i}(\lambda; \alpha_i, \beta_i) d\lambda = \\ &= \sum_{i=1}^n w_i m_{2,i} = \sum_{i=1}^n w_i(\sigma_i^2 + \mu_i^2) \\ \sigma^2 &= \mathbb{E}[\Lambda^2] - \mu^2 = \sum_{i=1}^n w_i(\sigma_i^2 + \mu_i^2) - \mu^2. \end{aligned}$$

This step is essential, as the variance of the mixture is not simply the average of the individual variances; instead, it requires the computation of the second moments. Under this result, taking the mean and variance of Gamma distribution into account, the shape and scale parameters are obtained by matching the first and second moments of the exact mixture. That is, estimated parameters are

$$\begin{aligned} \hat{\alpha} &= \mu^2/\sigma^2, \\ \hat{\beta} &= \sigma^2/\mu. \end{aligned}$$

This procedure, known as moment matching, projects the mixture distribution, which does not belong to the Gamma family except in special cases, onto the Gamma family while preserving its first two moments. From the perspective of mixture theory and the approximation of exponential families, this approach provides a tractable parametric representation. It is widely used in risk analysis, queuing theory, Bayesian inference, and the broader literature on mixture distributions when a simple parametric approximation is required [33]

This methodology has been implemented in R to facilitate its use in the scientific community.

### 2.3.2. Minimizing the Kullback–Leibler divergence

This second proposed methodology is based on minimizing the Kullback–Leibler divergence between two continuous probability distributions. This methodology was introduced in [34]. On the other hand, the KL divergence, including its definition, properties, and applications in statistics and inference is extensively covered in [35]. For each component, the mixture distribution of the Gamma distributions obtained for the failure rate is considered, whose density is denoted by $f$. Then, the parameters of a distribution with density $g$ that minimize the KL divergence between the two distributions are determined by maximum likelihood.

Formally, the problem is formulated as follows. Let the mixture

$$f(\lambda) = \sum_{i=1}^n w_i f_{\Lambda_i}(\lambda; \alpha_i, \beta_i), \quad \lambda > 0,$$

whose objective is to find the parameters that minimize

$$D_{KL}(f; |; g_\theta) = \int_0^\infty f(x) \log \frac{f(x)}{g_\theta(x)} dx.$$

Since the function $f$ does not depend on the parameter vector, this minimization is equivalent to maximizing

$$\mathcal{L}(\theta) = \int_0^\infty f(x) \log g_\theta(x) dx. \tag{1}$$

In the particular case where the approximating distribution is a Gamma distribution denoted as $\Lambda$ with density

$$g(\lambda; \alpha, \beta) = \frac{1}{\beta^\alpha \Gamma(\alpha)} \lambda^{\alpha-1} e^{-\lambda/\beta},$$

the expression (1) is rewritten as

$$\mathcal{L}(\alpha, \beta) = \int_0^\infty f(\lambda) \left[-\alpha \log \beta - \log \Gamma(\alpha) + (\alpha - 1) \log \lambda - \lambda/\beta\right] d\lambda. \tag{2}$$

Using the fact that $f$ is the probability density function of $\Lambda$, and therefore

$$\int_0^\infty f(\lambda)\, d\lambda = 1, \quad \mathbb{E}_f[\Lambda] = \int_0^\infty \lambda f(\lambda)\, d\lambda, \quad \mathbb{E}_f[\log \Lambda] = \int_0^\infty \log(\lambda) f(\lambda)\, d\lambda,$$

the expression (2) can be formulated as

$$\begin{aligned} \mathcal{L}(\alpha, \beta) &= -\alpha \log \beta \int_0^\infty f(\lambda)\, d\lambda - \log \Gamma(\alpha) \int_0^\infty f(\lambda)\, d\lambda \\ &\quad + (\alpha - 1) \int_0^\infty \log(\lambda) f(\lambda)\, d\lambda - \frac{1}{\beta} \int_0^\infty \lambda f(\lambda)\, d\lambda \\ &= -\alpha \log \beta - \log \Gamma(\alpha) + (\alpha - 1)\mathbb{E}_f[\log \Lambda] - \frac{\mathbb{E}_f[\Lambda]}{\beta}. \end{aligned} \tag{3}$$

Thus, once both expectations are obtained analytically or numerically, the optimization problem involves only two variables. If the function $f$ is the mixture of Gamma distributions, then

$$\begin{aligned} \mathbb{E}_f[\Lambda] &= \sum_{i=1}^n w_i \alpha_i \beta_i \\ \mathbb{E}_f[\log \Lambda] &= \sum_{i=1}^n w_i \mathbb{E}_{\Lambda_i}[\log \Lambda_i]. \end{aligned}$$

Clearly, if, in general, $X$ follows any Gamma distribution

$$\mathbb{E}_f[\log X] = \int_0^\infty \frac{\log x}{\beta^\alpha \Gamma(\alpha)} x^{\alpha-1} e^{-x/\beta} dx,$$

which is equivalent, after the change of variable $u = x/\beta$ (so that $x = \beta u$ and $dx = \beta\, du$), to

$$\begin{aligned} \mathbb{E}_f[\log X] &= \frac{1}{\Gamma(\alpha)} \int_0^\infty \log(u\beta) u^{\alpha-1} e^{-u} du \\ &= \frac{1}{\Gamma(\alpha)} \left( \int_0^\infty (\log u) u^{\alpha-1} e^{-u} du + \Gamma(\alpha) \log \beta \right) \\ &= \frac{1}{\Gamma(\alpha)} \int_0^\infty (\log u) u^{\alpha-1} e^{-u} du + \log \beta. \end{aligned}$$

Then, $\mathbb{E}_f[\log X] = \psi(\alpha) + \log \beta$, with $\psi$ being the digamma function $\psi(\alpha) = \frac{d \log \Gamma(\alpha)}{d\alpha}$.

Returning to the problem considered in this research, this yields

$$\mathbb{E}_f[\log \Lambda] = \sum_{i=1}^n w_i \left( \psi(\alpha_i) + \log \beta_i \right).$$

To optimize the objective, the following expressions are derived from (3)

$$\begin{aligned} \frac{\partial \mathcal{L}(\alpha, \beta)}{\partial \beta} &= -\frac{\alpha}{\beta} + \frac{\mathbb{E}_f[\Lambda]}{\beta^2} = 0 \\ \frac{\partial \mathcal{L}(\alpha, \beta)}{\partial \alpha} &= -\log \beta - \psi(\alpha) + \mathbb{E}_f[\log \Lambda] = 0. \end{aligned}$$

Hence,

$$-\log \left( \frac{\mathbb{E}_f[\Lambda]}{\hat{\alpha}} \right) - \psi(\hat{\alpha}) + \mathbb{E}_f[\log \Lambda] = 0, \tag{4}$$

and

$$\hat{\beta} = \frac{\mathbb{E}_f[\Lambda]}{\hat{\alpha}}.$$

The Eq. (4) is a nonlinear equation that can be solved numerically, yielding the desired maximization. As a convenient starting value for the numerical solution, the method of moments estimator derived in the previous section can be used as follows:

$$\alpha_0 = \frac{\left(\mathbb{E}_f[\Lambda]\right)^2}{\mathrm{Var}_f[\Lambda]}.$$

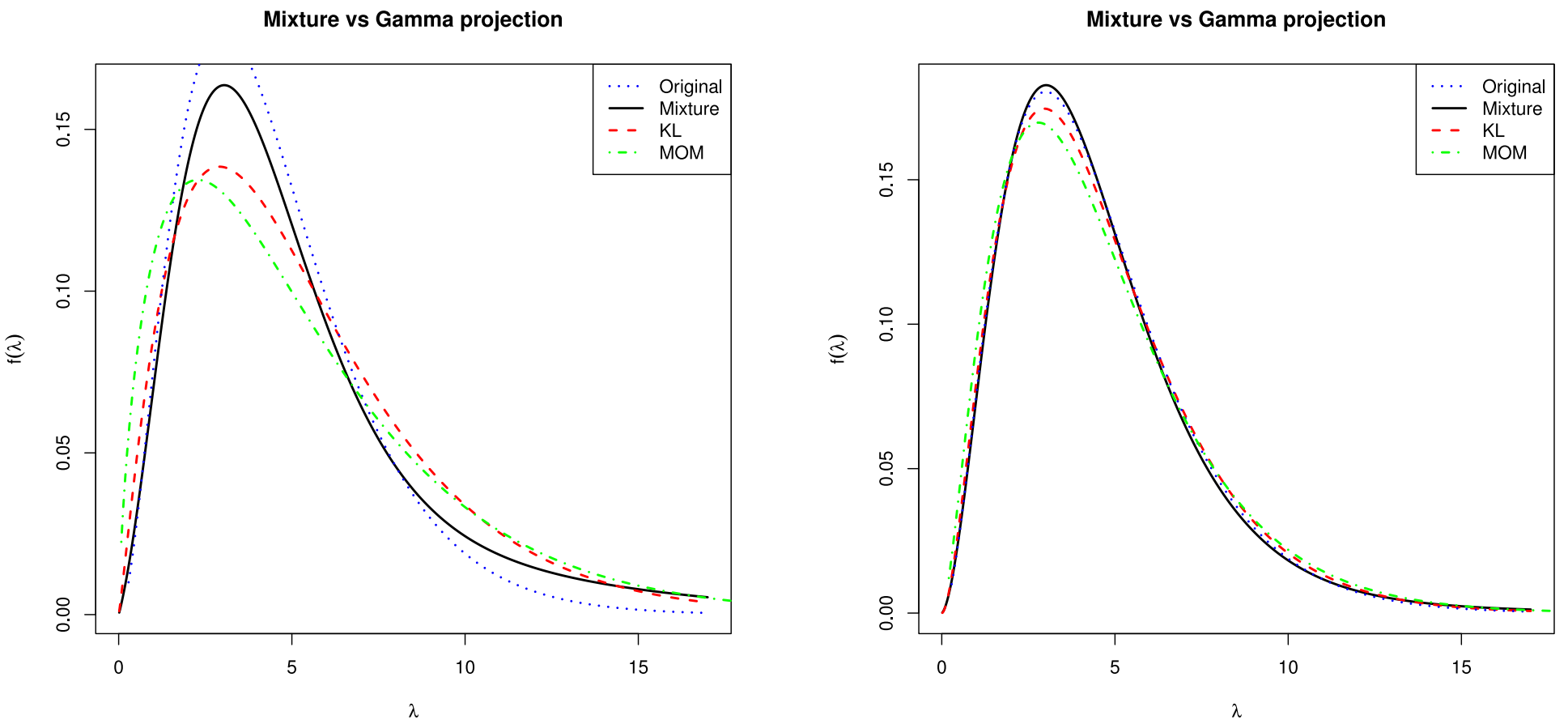

**Fig. 1.** Probability density functions for the corresponding distributions in the generic simulation case: original, mixture, moment and KL methodologies for $n = 5$ (left) and $n = 50$ (right) under the heterogeneity scenario.

**Table 1**
Results of simulation $\Gamma(\alpha = 3, \beta = 1.5)$ with different number of studies by moment matching and KL divergence.

| Studies | Moments | | KL divergence | |
|---|---|---|---|---|
| | Homogeneity | Heterogeneity | Homogeneity | Heterogeneity |
| $n = 5$ | $\hat{\alpha} = 3.029, \hat{\beta} = 1.494$ | $\hat{\alpha} = 1.709, \hat{\beta} = 3.147$ | $\hat{\alpha} = 3.044, \hat{\beta} = 1.487$ | $\hat{\alpha} = 2.143, \hat{\beta} = 2.511$ |
| $n = 50$ | $\hat{\alpha} = 2.987, \hat{\beta} = 1.507$ | $\hat{\alpha} = 2.551, \hat{\beta} = 1.816$ | $\hat{\alpha} = 3.001, \hat{\beta} = 1.501$ | $\hat{\alpha} = 2.795, \hat{\beta} = 1.658$ |
| $n = 100$ | $\hat{\alpha} = 2.989, \hat{\beta} = 1.505$ | $\hat{\alpha} = 2.738, \hat{\beta} = 1.668$ | $\hat{\alpha} = 3.003, \hat{\beta} = 1.498$ | $\hat{\alpha} = 2.892, \hat{\beta} = 1.580$ |
| $n = 500$ | $\hat{\alpha} = 2.985, \hat{\beta} = 1.507$ | $\hat{\alpha} = 2.928, \hat{\beta} = 1.541$ | $\hat{\alpha} = 2.999, \hat{\beta} = 1.500$ | $\hat{\alpha} = 2.975, \hat{\beta} = 1.517$ |

In this sense, it is important to emphasize that the final KL estimates are determined by the solution of Eq. (4) and, therefore, are independent of the method of moments estimates once convergence is achieved.

This methodology has been implemented in R, using both the numerical bisection method and the well-known Newton–Raphson algorithm to find the optimum.

## 3. Simulation

To assess the performance of the proposed methodology to integrate multi-source heterogeneous data for the same component, a simulation study is conducted.

Let us consider a target Gamma distribution, known a priori, which serves as a benchmark for comparison with the distribution obtained through this methodology. In particular, let be $\Lambda \sim \Gamma(\alpha, \beta)$ a random variable where $\alpha$ and $\beta$ are the shape and scale parameters, respectively. Then, $n$ samples are simulated from this distribution $\Gamma(\alpha, \beta)$, representing each of the studies with sample size $N_i$; $i = 1, \ldots, n$, which is randomly chosen between 50 and 100. Then, the corresponding Gamma distributions are fitted by maximum likelihood for each sample, that is, $\Lambda_i \sim \Gamma(\alpha_i, \beta_i)$ with $i = 1, \ldots, n$. In addition, different sample sizes $n = 5, 10, 50, 100, 500$ are considered to evaluate the effect of the number of studies in the estimation process. Once the parameters of $\Lambda_i$ are estimated, the method of moments and the methodology based on minimizing the Kullback–Leibler divergence are applied. Here, it is assumed that all the studies have the same importance, i.e., $w_1 = \cdots = w_n$. In this step, the original distribution parameters are achieved, i.e., $\hat{\alpha}$ and $\hat{\beta}$. For each $n$, the analysis is repeated 100 times and the mean of these 100 replications are taken as parameter estimates.

Table 1 shows the obtained results for the distribution $\Gamma(\alpha = 3, \beta = 1.5)$. Furthermore, to evaluate the robustness of the proposed methodology under heterogeneous conditions, additional Gamma distributions with different parameter values are incorporated into the simulation. These distributions are intended to emulate the heterogeneity among studies for the same component that may arise in practice due to different sources of variation. In particular, the following distributions are considered: $Y_1 \sim \Gamma(\alpha = 4, \beta = 1)$, $Y_2 \sim \Gamma(\alpha = 2, \beta = 2)$ and $Y_3 \sim \Gamma(\alpha = 5, \beta = 2.5)$, which are added to $\Lambda_i$ in the methodology to estimate the original distribution. Besides, Figs. 1 and 2 provide a comparative view, based on the simulation, of the original density, the mixture and the Gamma distribution obtained using the method of moments and the KL-based method for the heterogeneity case. Finally, Fig. 3 illustrates the uncertainty associated with the estimation of parameters for each case.

A similar study was conducted on the interface flange component reported in Table 1 of [14], with the aim of evaluating the proposed methodology using data representative of those typically employed in DONES Accelerator Systems (here, the variable of interest $\Lambda$ corresponds to the failure rate, expressed in failures per hour). In particular, the reference distribution $\Gamma(\alpha, \beta)$ is defined by the arithmetic mean of the estimated shape and scale parameters, $\alpha_i$ and $\beta_i$, yielding $\Gamma(\alpha = 0.718, \beta = 3.813 \times 10^{-6})$. Again, $n$ samples are subsequently generated from this distribution, and the corresponding results are reported in Table 2. To assess the impact of heterogeneity, the simulation is deliberately perturbed by incorporating parameter estimates from the three cases reported in Table 1 of [14]. Specifically, the distributions $Y_1 \sim \Gamma(\alpha = 0.242, \beta = 1.241 \times 10^{-6})$, $Y_2 \sim \Gamma(\alpha = 0.229, \beta = 1.128 \times 10^{-5})$, and $Y_3 \sim \Gamma(\alpha = 1.699, \beta = 3.533 \times 10^{-8})$ are incorporated into the set $\Lambda_i$ within the proposed methodology to estimate the original underlying distribution. It should be noted that these Gamma distributions were obtained by applying the methodology described in Section 2.2, using the initial information presented in the Table 1 of [14]. In addition, Figs. 4 and 5 present a comparative, simulation-based illustration of the original density, the mixture distribution, and the Gamma distributions obtained using the method of moments and the Kullback–Leibler-based approach for the heterogeneous case. Finally, Fig. 6 presents empirical distribution of the parameters estimates under the different simulation settings.

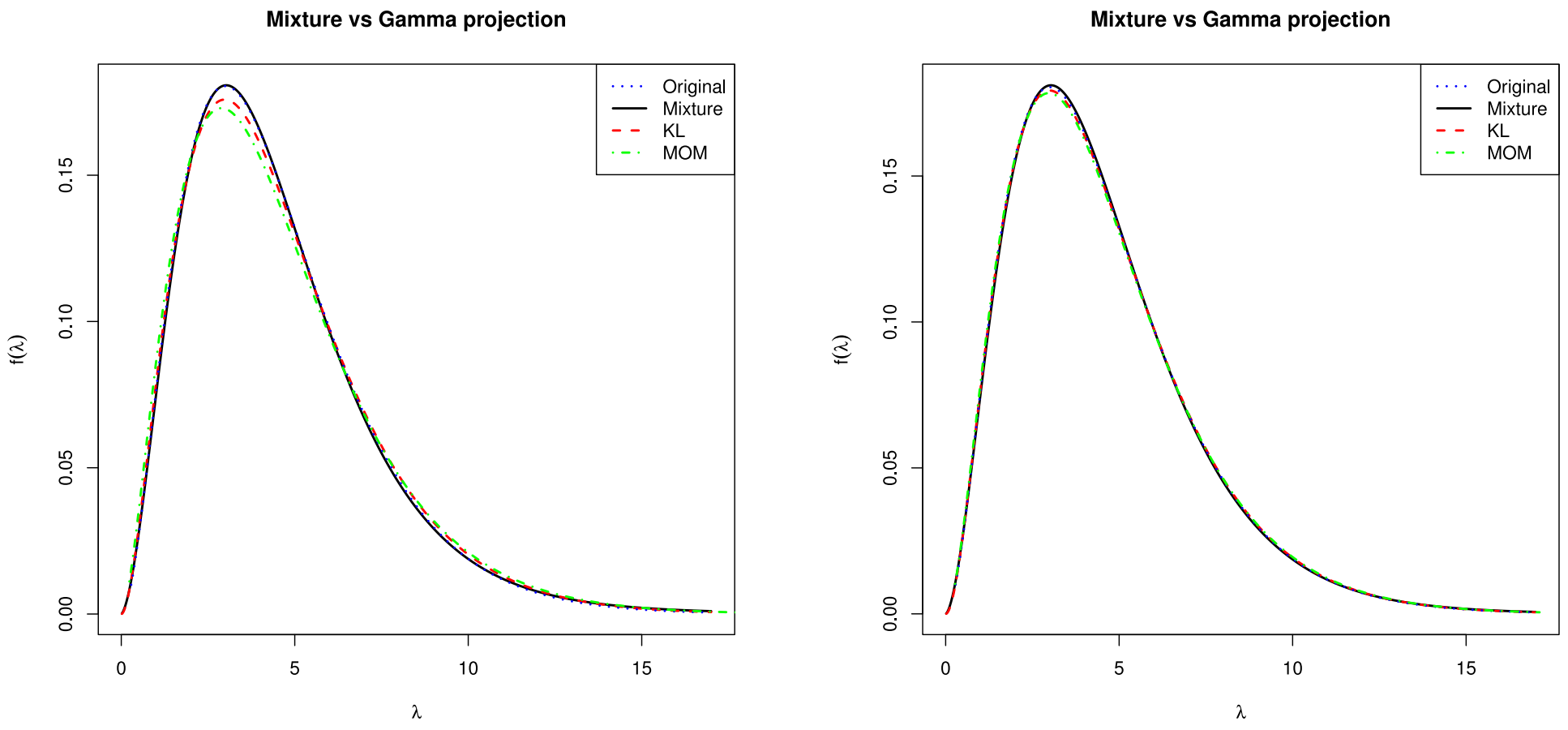

**Fig. 2.** Probability density functions for the corresponding distributions in the generic simulation case: original, mixture, moment and KL methodologies for $n = 100$ (left) and $n = 500$ (right) under the heterogeneity scenario.

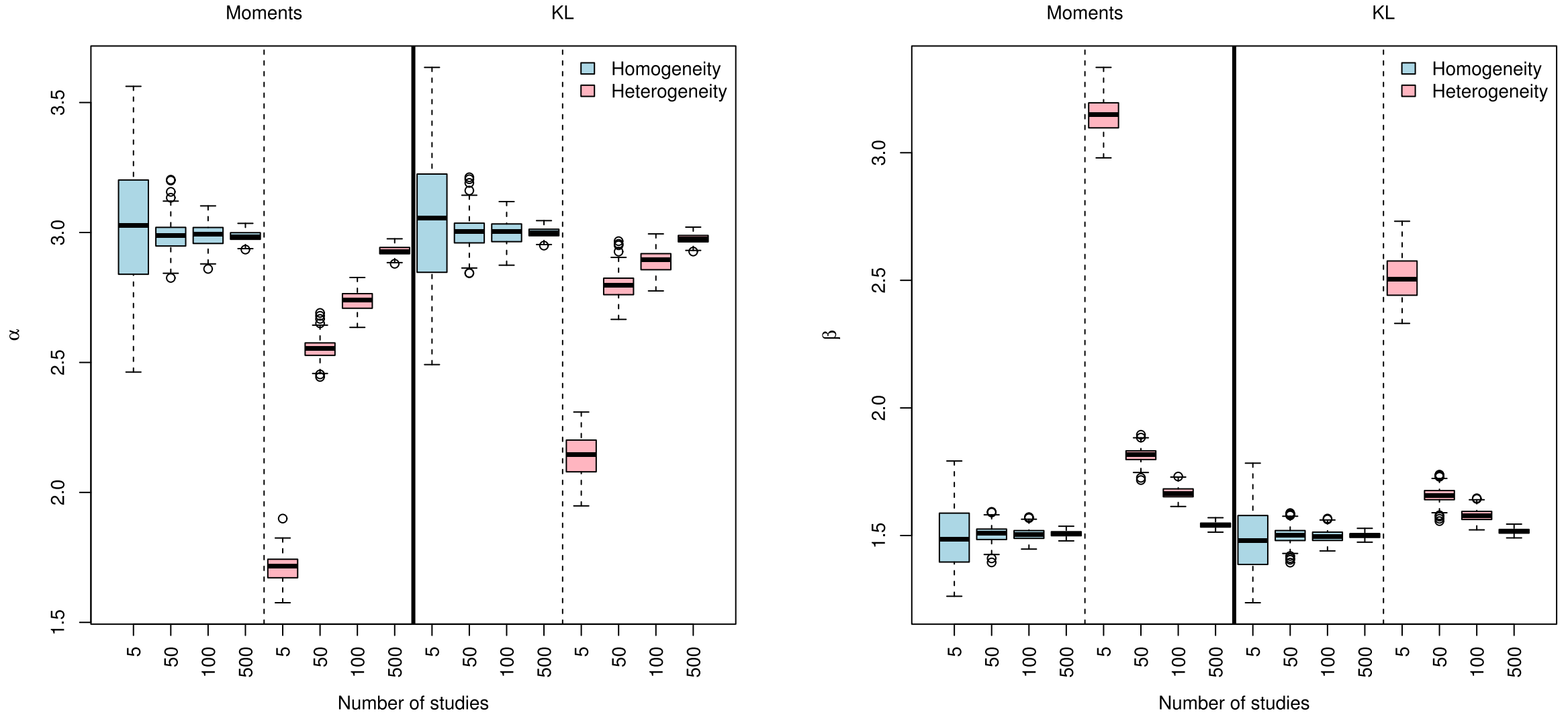

**Fig. 3.** Boxplots of the estimated values of $\alpha$ (left) and $\beta$ (right) across sample sizes and estimation methods, under homogeneous and heterogeneous scenarios in the generic simulation case.

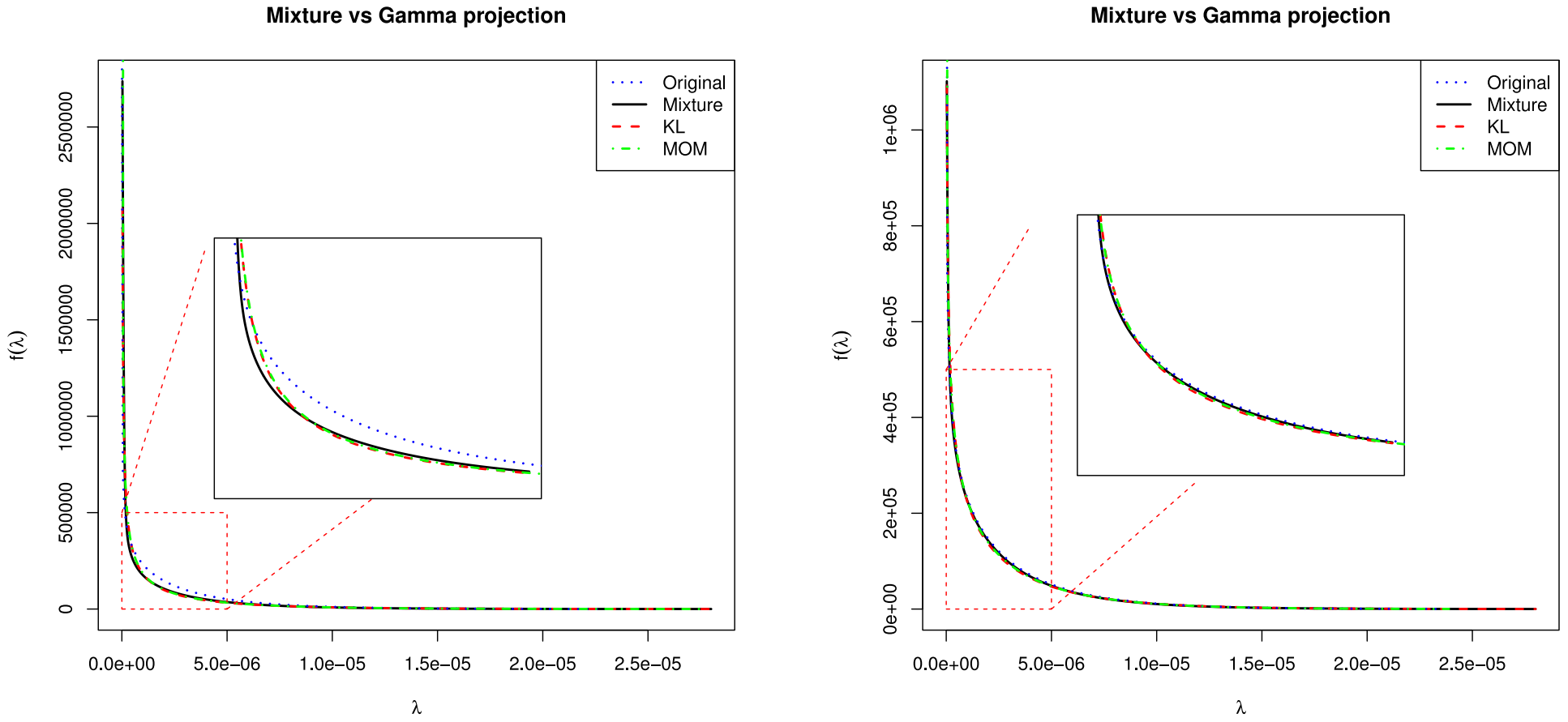

**Fig. 4.** Probability density functions for the corresponding distributions in the interface flange component simulation case: original, mixture, moment and KL methodologies for $n = 5$ (left) and $n = 50$ (right) under the heterogeneity scenario. Insets display an enlarged view of the region $x \in [0, 5 \times 10^{-6}]$ and $y \in [0, 500000]$ to facilitate comparison of the distributions in this region.

**Table 2**
Results of simulation $\Gamma(\alpha = 0.718, \beta = 3.813 \times 10^{-6})$ with different number of studies by moment matching and KL divergence.

| Studies | Moments | | KL divergence | |
|---|---|---|---|---|
| | Homogeneity | Heterogeneity | Homogeneity | Heterogeneity |
| $n = 5$ | $\hat{\alpha} = 0.707, \hat{\beta} = 3.873 \times 10^{-6}$ | $\hat{\alpha} = 0.373, \hat{\beta} = 5.552 \times 10^{-6}$ | $\hat{\alpha} = 0.718, \hat{\beta} = 3.811 \times 10^{-6}$ | $\hat{\alpha} = 0.368, \hat{\beta} = 5.638 \times 10^{-6}$ |
| $n = 50$ | $\hat{\alpha} = 0.705, \hat{\beta} = 3.882 \times 10^{-6}$ | $\hat{\alpha} = 0.643, \hat{\beta} = 4.110 \times 10^{-6}$ | $\hat{\alpha} = 0.718, \hat{\beta} = 3.808 \times 10^{-6}$ | $\hat{\alpha} = 0.621, \hat{\beta} = 4.244 \times 10^{-6}$ |
| $n = 100$ | $\hat{\alpha} = 0.705, \hat{\beta} = 3.887 \times 10^{-6}$ | $\hat{\alpha} = 0.672, \hat{\beta} = 3.999 \times 10^{-6}$ | $\hat{\alpha} = 0.719, \hat{\beta} = 3.809 \times 10^{-6}$ | $\hat{\alpha} = 0.665, \hat{\beta} = 4.043 \times 10^{-6}$ |
| $n = 500$ | $\hat{\alpha} = 0.703, \hat{\beta} = 3.891 \times 10^{-6}$ | $\hat{\alpha} = 0.696, \hat{\beta} = 3.913 \times 10^{-6}$ | $\hat{\alpha} = 0.718, \hat{\beta} = 3.811 \times 10^{-6}$ | $\hat{\alpha} = 0.706, \hat{\beta} = 3.860 \times 10^{-6}$ |

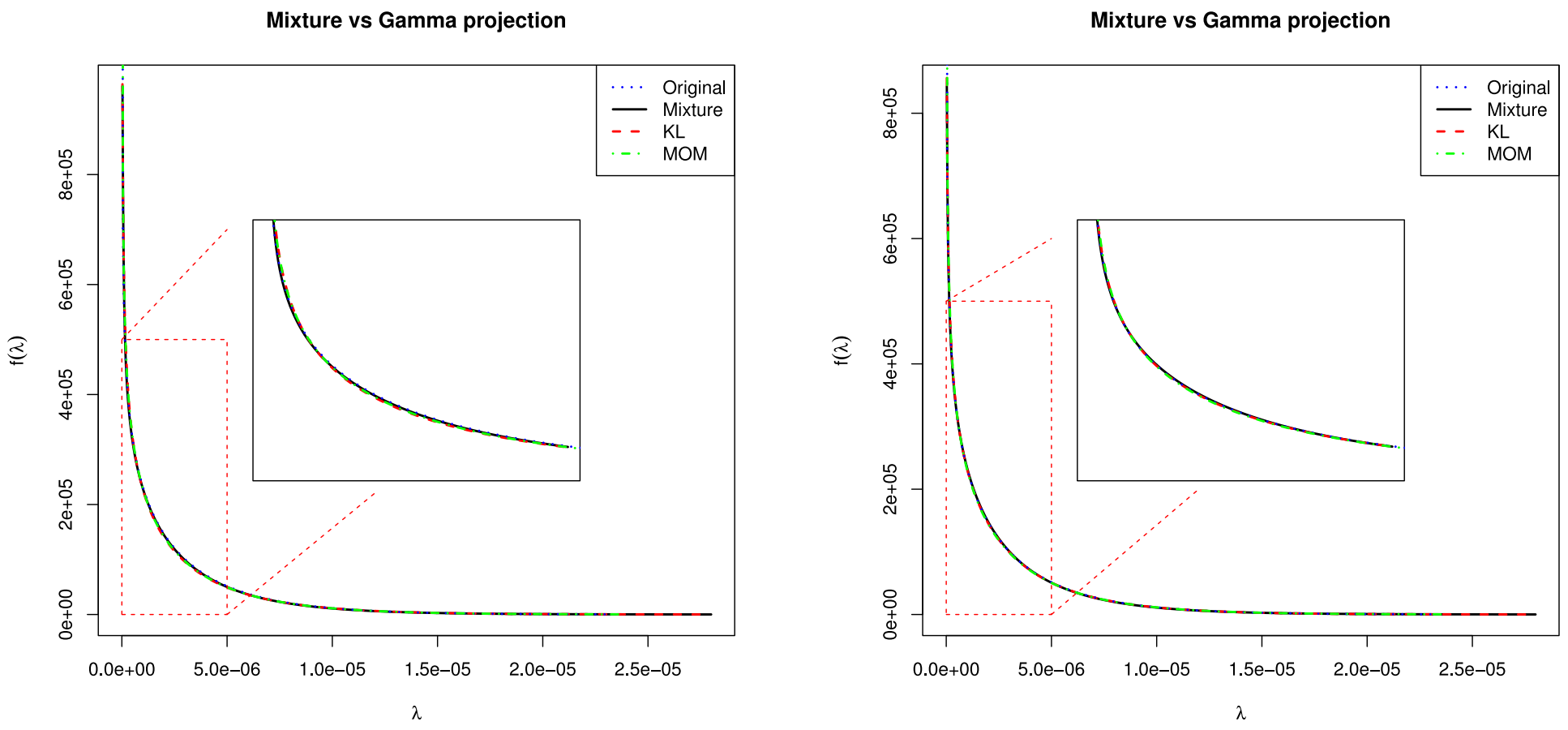

**Fig. 5.** Probability density functions for the corresponding distributions in the interface flange component simulation case: original, mixture, moment and KL methodologies for $n = 100$ (left) and $n = 500$ (right) under the heterogeneity scenario. Insets display an enlarged view of the region $x \in [0, 5 \times 10^{-6}]$ and $y \in [0, 500000]$ to facilitate comparison of the distributions in this region.

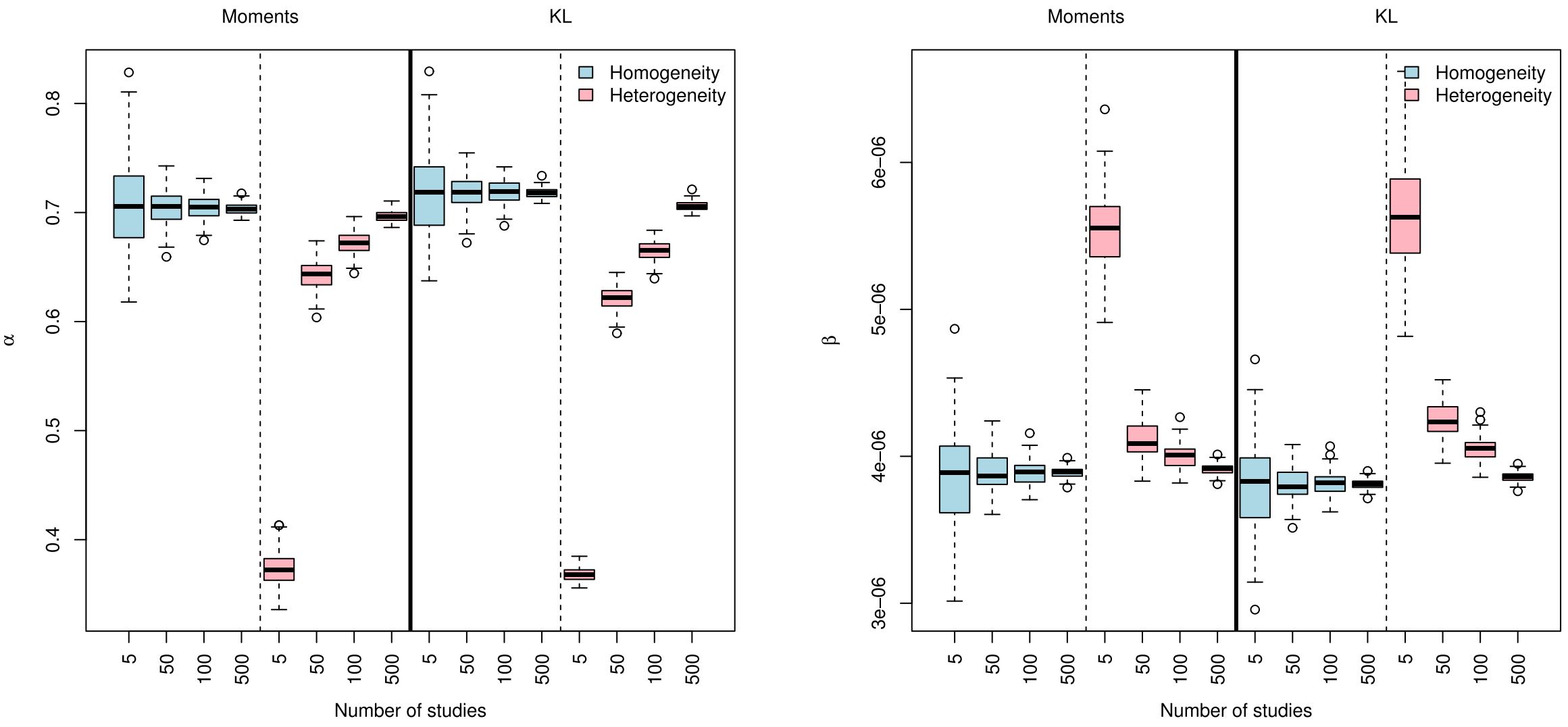

**Fig. 6.** Boxplots of the estimated values of $\alpha$ (left) and $\beta$ (right) across sample sizes and estimation methods, under homogeneous and heterogeneous scenarios in the interface flange component simulation case.

The results presented above indicate that the proposed methodology yields very promising outcomes. As shown by the boxplots, the proposed methodologies exhibit relatively low variability in the parameter estimates across the different sample sizes. Although a slightly higher degree of variability can be observed for the smallest sample size ($n = 5$), the overall uncertainty remains limited. In relation to the level of precision, the number of studies has a relevant impact only when heterogeneity exists among data sources, in which the KL divergence approach provides more accurate estimates. Conversely, under homogeneous conditions, both integration methods achieve a high level of accuracy in recovering the original distribution, even with a relatively limited number of studies. In this setting, the method of moments provides comparable parameter estimates while requiring substantially less computational effort than the KL divergence approach, which requires solving an optimization problem involving a nonlinear equation numerically. Therefore, the method of moments can be considered a suitable and computationally efficient alternative under homogeneous conditions, whereas the KL divergence methodology is preferred when heterogeneity among data sources is present.

**Table 3**
Estimates for the interface flange component. All failure rates are expressed in failures per hour (/h).

| Failure Mode | Probability Distribution Function (PDF) | $\lambda_{mean}$ (Ref.Mean) | $\lambda_{median}$ $(Md)$ | $\lambda_{5\%}$ $(LB)$ | $\lambda_{95\%}$ $(UB)$ | Error Factor $(EF)$ | # |
|---|---|---|---|---|---|---|---|
| Leakage | Unspecified | 2.60e−07/h | | | | 5 | 1 |
| Leakage | Lognormal(−1.61e+01,1.4e+00) | 2.60e−07/h | 9.76e−08/h | 9.76e−09/h | 9.76e−07/h | 10 | 1 |
| Leakage | Gamma(8.33e−01,3.12e−07) | 2.60e−07/h | 1.66e−07/h | 8.06e−09/h | 8.31e−07/h | 5 | 1 |
| Leakage | Unspecified | 5.70e−07/h | | | | 5 | 2 |
| Leakage | Lognormal(−1.54e+01,1.4e+00) | 5.70e−07/h | 2.14e−07/h | 2.14e−08/h | 2.14e−06/h | 10 | 2 |
| Leakage | Gamma(8.33e−01,6.84e−07) | 5.70e−07/h | 3.64e−07/h | 1.77e−08/h | 1.82e−06/h | 5 | 2 |
| Leakage | Unspecified | 9.80e−07/h | | | | 5 | 3 |
| Leakage | Lognormal(−1.48e+01,1.4e+00) | 9.80e−07/h | 3.68e−07/h | 3.68e−08/h | 3.68e−06/h | 10 | 3 |
| Leakage | Gamma(8.33e−01,1.18e−06) | 9.80e−07/h | 6.27e−07/h | 3.04e−08/h | 3.13e−06/h | 5 | 3 |
| Leakage | Unspecified | 2.80e−06/h | | | | 5 | 4 |
| Leakage | Lognormal(−1.38e+01,1.4e+00) | 2.80e−06/h | 1.05e−06/h | 1.05e−07/h | 1.05e−05/h | 10 | 4 |
| Leakage | Gamma(8.33e−01,3.36e−06) | 2.80e−06/h | 1.79e−06/h | 8.68e−08/h | 8.95e−06/h | 5 | 4 |
| Leakage – Rupture | Lognormal | 2.58e−06/h | 3.00e−07/h | 1.00e−08/h | 1.00e−05/h | | 5 |
| Leakage – Rupture | Lognormal(−1.5e+01,2.1e+00) | 2.87e−06/h | 3.16e−07/h | 1.00e−08/h | 1.00e−05/h | 31.6 | 5 |
| Leakage – Rupture | Gamma(2.29e−01,1.13e−05) | 2.58e−06/h | 3.72e−07/h | 1.53e−11/h | 1.28e−05/h | 34.4 | 5 |
| Leakage – Rupture | Lognormal | 3.00e−06/h | | | | 30 | 6 |
| Leakage – Rupture | Lognormal(−1.49e+01,2.07e+00) | 3.00e−06/h | 3.54e−07/h | 1.18e−08/h | 1.06e−05/h | 30 | 6 |
| Leakage – Rupture | Gamma(2.42e−01,1.24e−05) | 3.00e−06/h | 4.89e−07/h | 3.46e−11/h | 1.47e−05/h | 30 | 6 |
| Leakage – Rupture | Lognormal | 3.00e−07/h | | | | 30 | 7 |
| Leakage – Rupture | Lognormal(−1.72e+01,2.07e+00) | 3.00e−07/h | 3.54e−08/h | 1.18e−09/h | 1.06e−06/h | 30 | 7 |
| Leakage – Rupture | Gamma(2.42e−01,1.24e−06) | 3.00e−07/h | 4.89e−08/h | 3.46e−12/h | 1.47e−06/h | 30 | 7 |
| Leakage | Exponential | 6.00e−08/h | | | 1.50e−07/h | | 8 |
| Leakage | Lognormal(−1.78e+01,1.28e+00) | 4.15e−08/h | 1.84e−08/h | 2.25e−09/h | 1.50e−07/h | 8.2 | 8 |
| Leakage | Gamma(1.70e+00,3.53e−08) | 6.00e−08/h | 4.87e−08/h | 8.54e−09/h | 1.50e−07/h | 3.1 | 8 |

**Table 4**
Estimates for the vacuum pump component. All failure rates are expressed in failures per hour (/h).

| Failure Mode | Probability Distribution Function (PDF) | $\lambda_{mean}$ (Ref.Mean) | $\lambda_{median}$ $(Md)$ | $\lambda_{5\%}$ $(LB)$ | $\lambda_{95\%}$ $(UB)$ | Error Factor $(EF)$ | # |
|---|---|---|---|---|---|---|---|
| All failure modes | Exponential | 3.80e−05/h | | 3e−05/h | 4.80e−05/h | | 9 |
| All failure modes | Lognormal(−1.02e+01,1.43e−01) | 3.83e−05/h | 3.79e−05/h | 3.00e−05/h | 4.80e−05/h | 1.3 | 9 |
| All failure modes | Gamma(4.94e+01,7.69e−07) | 3.80e−05/h | 3.77e−05/h | 2.96e−05/h | 4.73e−05/h | 1.3 | 9 |
| All failure modes | Unspecified | 3.20e−04/h | | | | 5 | 10 |
| All failure modes | Lognormal(−9.03e+00,1.4e+00) | 3.20e−04/h | 1.20e−04/h | 1.20e−05/h | 1.20e−03/h | 10 | 10 |
| All failure modes | Gamma(8.33e−01,3.84e−04) | 3.20e−04/h | 2.05e−04/h | 9.92e−06/h | 1.02e−03/h | 5 | 10 |

## 4. Application with real data

The methodology described is applied to an empirical dataset for selected equipment of the IFMIF-DONES accelerator systems. Specifically, the data reported in Table 1 of [14] are considered, covering failure statistics of the equipment operated in the fusion-oriented facilities and reported in FCFRDB database [7], including interface flanges, vacuum pumps, high voltage transformers, radio frequency amplifiers, helium compressors, and pressure sensors. The exemplary components were selected based on two criteria: (1) uniqueness to fusion applications, thus excluding commonly discussed in the literature components, such as pumps, valves, bearings, (2) availability of a substantial number of data sources, thus demonstrating the need for the improved data aggregation algorithm.

As a first step, this table is re-estimated by using the Bayesian result given in Section 2.2, and the resulting estimates are compared with those reported in the original study. To facilitate the comparison, the results for each component are presented separately (see Tables 3, 4, 5, 6, 7 and 8). In each table, the first row (white) corresponds to the original data from the database, the second row (gray) reports the estimates obtained by the authors of the referenced article, and the third row (blue) presents the estimates obtained using the proposed Gamma distribution. Parameters assumed to be fixed are highlighted in yellow. Note that empty cells indicate information that is not provided in the original database.

It can be observed that the reference mean failure rate (the main metric used to evaluate and characterize the frequency of failures, as previously mentioned) matches in all cases the values obtained using the proposed Gamma methodology. Consequently, the proposed methodology not only improves the precision and rigor of the results but is also supported by a solid Bayesian theoretical methodology.

Once the failure rate distributions derived from heterogeneous and incomplete data have been estimated, the next step is to integrate the multiple probability density functions corresponding to the same component into a unified density that captures the variability of the failure rate for each component. Due to the limited number of available studies for each component in the dataset, no distinction was made in this application between the different failure modes of each component. This assumption allows the available information to be integrated within the proposed framework. However, distinguishing between failure modes should be explicitly considered in future applications whenever sufficient data are available, in order to ensure the engineering validity of the results.

To this end, the moments-based and Kullback–Leibler divergence methodologies described in Section 2.3 are employed, whose resulting estimates are reported in Table 9. These unified distributions constitute the failure rate inputs required by the optimal maintenance planning framework of IFMIF-DONES; their practical role within that framework is discussed in Section 5.

## 5. Discussion and open lines

The methodology proposed in this manuscript is motivated by dealing with multiple information sources that differ in structure and content, generated by a range of organizations involved in both fusion and fission facility operations in the context of fusion-specific components of the DONES Accelerator Systems.

Firstly, to address the problem of estimating the probability density of the failure rate from multiple databases for the same component

**Table 5**

Estimates for the RF amplifier component. All failure rates are expressed in failures per hour (/h).

| Failure Mode | Probability Distribution Function (PDF) | $\lambda_{mean}$ (Ref.Mean) | $\lambda_{median}$ ($Md$) | $\lambda_{5\%}$ ($LB$) | $\lambda_{95\%}$ ($UB$) | Error Factor ($EF$) | # |
|---|---|---|---|---|---|---|---|
| All failure modes | Exponential | 4.50e−06/h | | 2.70e−06/h | 7e−06/h | | 11 |
| All failure modes | Lognormal(−1.23e+01,2.9e−01) | 4.53e−06/h | 4.35e−06/h | 2.70e−06/h | 7.00e−06/h | 1.6 | 11 |
| All failure modes | Gamma(1.23e+01,3.64e−07) | 4.50e−06/h | 4.38e−06/h | 2.62e−06/h | 6.79e−06/h | 1.6 | 11 |
| Erratic/No output | Exponential | 5.50e−06/h | | 3.50e−06/h | 8.20e−06/h | | 12 |
| Erratic/No output | Lognormal(−1.21e+01,2.59e−01) | 5.54e−06/h | 5.36e−06/h | 3.50e−06/h | 8.20e−06/h | 1.5 | 12 |
| Erratic/No output | Gamma(1.54e+01,3.58e−07) | 5.50e−06/h | 5.38e−06/h | 3.41e−06/h | 7.99e−06/h | 1.5 | 12 |
| All failure modes | Exponential | 2.68e−03/h | | | | 5 | 13 |
| All failure modes | Lognormal(−6.9e+00,1.4e+00) | 2.68e−03/h | 1.01e−03/h | 1.01e−04/h | 1.01e−02/h | 10 | 13 |
| All failure modes | Gamma(8.33e−01,3.22e−03) | 2.68e−03/h | 1.71e−03/h | 8.31e−05/h | 8.57e−03/h | 5 | 13 |
| All failure modes | Exponential | 4.92e−02/h | | | | 5 | 14 |
| All failure modes | Lognormal(−4.13e+00,1.4e+00) | 4.29e−02/h | 1.61e−02/h | 1.61e−03/h | 1.61e−01/h | 10 | 14 |
| All failure modes | Gamma(8.33e−01,5.91e−02) | 4.92e−02/h | 3.15e−02/h | 1.53e−03/h | 1.57e−01/h | 5 | 14 |

**Table 6**

Estimates for the HV transformer component. All failure rates are expressed in failures per hour (/h).

| Failure Mode | Probability Distribution Function (PDF) | $\lambda_{mean}$ (Ref.Mean) | $\lambda_{median}$ ($Md$) | $\lambda_{5\%}$ ($LB$) | $\lambda_{95\%}$ ($UB$) | Error Factor ($EF$) | # |
|---|---|---|---|---|---|---|---|
| All failure modes | Unspecified | 9.60e−06/h | | | | 5 | 15 |
| All failure modes | Lognormal(−1.25e+01,1.4e+00) | 9.60e−06/h | 3.60e−06/h | 3.60e−07/h | 3.60e−05/h | 10 | 15 |
| All failure modes | Gamma(8.33e−01,1.15e−05) | 9.60e−06/h | 6.14e−06/h | 2.98e−07/h | 3.07e−05/h | 5 | 15 |
| All failure modes | Exponential | 2.10e−05/h | | 1.6e−05/h | 2.60e−05/h | | 16 |
| All failure modes | Lognormal(−1.08e+01,1.48e−01) | 2.06e−05/h | 2.04e−05/h | 1.60e−05/h | 2.60e−05/h | 1.3 | 16 |
| All failure modes | Gamma(4.63e+01,4.53e−07) | 2.10e−05/h | 2.08e−05/h | 1.62e−05/h | 2.63e−05/h | 1.3 | 16 |
| All failure modes | Exponential | 3.10e−04/h | | 2.40e−04/h | 3.80e−04/h | | 17 |
| All failure modes | Lognormal(−8.11e+00,1.4e−01) | 3.05e−04/h | 3.02e−04/h | 2.40e−04/h | 3.80e−04/h | 1.3 | 17 |
| All failure modes | Gamma(5.17e+01,6.00e−06) | 3.10e−04/h | 3.08e−04/h | 2.43e−04/h | 3.84e−04/h | 1.2 | 17 |
| All failure modes | Exponential | 7.30e−06/h | | 4.5e−06/h | 1.10e−05/h | | 18 |
| All failure modes | Lognormal(−1.19e+01,2.72e−01) | 7.30e−06/h | 7.04e−06/h | 4.50e−06/h | 1.10e−05/h | 1.6 | 18 |
| All failure modes | Gamma(1.40e+01,5.23e−07) | 7.30e−06/h | 7.13e−06/h | 4.41e−06/h | 1.08e−05/h | 1.5 | 18 |
| All failure modes | Exponential | 7.90e−04/h | | 4.90e−04/h | 1.20e−03/h | | 19 |
| All failure modes | Lognormal(−7.17e+00,2.72e−01) | 7.96e−04/h | 7.67e−04/h | 4.90e−04/h | 1.20e−03/h | 1.6 | 19 |
| All failure modes | Gamma(1.39e+01,5.68e−05) | 7.90e−04/h | 7.71e−04/h | 4.77e−04/h | 1.17e−03/h | 1.5 | 19 |

**Table 7**

Estimates for the helium compressor component. All failure rates are expressed in failures on demand (/d).

| Failure Mode | Probability Distribution Function (PDF) | $\lambda_{mean}$ (Ref.Mean) | $\lambda_{median}$ ($Md$) | $\lambda_{5\%}$ ($LB$) | $\lambda_{95\%}$ ($UB$) | Error Factor ($EF$) | # |
|---|---|---|---|---|---|---|---|
| Failure to funct./oper. | Exponential | 2.75e−02/d | | 1.78e−02/d | 4.08e−02/d | | 20 |
| Failure to funct./oper. | Lognormal(−3.61e+00,2.52e−01) | 2.78e−02/d | 2.69e−02/d | 1.78e−02/d | 4.08e−02/d | 1.5 | 20 |
| Failure to funct./oper. | Gamma(1.62e+01,1.70e−03) | 2.75e−02/d | 2.69e−02/d | 1.73e−02/d | 3.96e−02/d | 1.5 | 20 |
| Failure to funct./oper. | Exponential | 9.17e−03/d | | 4.00e−03/d | 1.81e−02/d | | 21 |
| Failure to funct./oper. | Lognormal(−4.77e+00,4.59e−01) | 9.45e−03/d | 8.51e−03/d | 4.00e−03/d | 1.81e−02/d | 2.1 | 21 |
| Failure to funct./oper. | Gamma(5.16e+00,1.78e−03) | 9.17e−03/d | 8.59e−03/d | 3.68e−03/d | 1.67e−02/d | 1.9 | 21 |
| Failure to funct./oper. | Exponential | 3.97e−02/d | | 2.29e−02/d | 6.44e−02/d | | 22 |
| Failure to funct./oper. | Lognormal(−3.26e+00,3.14e−01) | 4.03e−02/d | 3.84e−02/d | 2.29e−02/d | 6.44e−02/d | 1.7 | 22 |
| Failure to funct./oper. | Gamma(1.05e+01,3.77e−03) | 3.97e−02/d | 3.85e−02/d | 2.19e−02/d | 6.17e−02/d | 1.6 | 22 |
| Failure to funct./oper. | Exponential | 6.62e−03/d | | 1.18e−03/d | 2.08e−02/d | | 23 |
| Failure to funct./oper. | Lognormal(−5.31e+00,8.72e−01) | 7.25e−03/d | 4.95e−03/d | 1.18e−03/d | 2.08e−02/d | 4.2 | 23 |
| Failure to funct./oper. | Gamma(1.70e+00,3.91e−03) | 6.62e−03/d | 5.37e−03/d | 9.39e−04/d | 1.66e−02/d | 3.1 | 23 |
| Fails to start (norm.run.) | Beta | 3.41e−02/d | 1.72e−02/d | 2.06e−04/d | 1.25e−01/d | 7.3 | 24 |
| Fails to start (norm.run.) | Beta(5.39e−01,1.53e+01) | 3.40e−02/d | 1.72e−02/d | 2.06e−04/d | 1.25e−01/d | 7.3 | 24 |
| Fails to start (norm.run.) | Gamma(5.69e−01,5.99e−02) | 3.41e−02/d | 1.73e−02/d | 2.53e−04/d | 1.25e−01/d | 7.3 | 24 |
| Fails to start (standby) | Beta | 4.16e−03/d | 3.14e−03/d | 3.58e−04/d | 1.15e−02/d | 3.7 | 25 |
| Fails to start (standby) | Beta(1.26e+00,3.01e+02) | 4.17e−03/d | 3.14e−03/d | 3.58e−04/d | 1.15e−02/d | 3.7 | 25 |
| Fails to start (standby) | Gamma(1.27e+00,3.27e−03) | 4.16e−03/d | 3.13e−03/d | 3.62e−04/d | 1.15e−02/d | 3.7 | 25 |
| Fails to start (standby) | Beta | 8.24e−03/d | 8.07e−03/d | 5.29e−03/d | 1.17e−02/d | 1.4 | 26 |
| Fails to start (standby) | Beta(1.75e+01,2.1e+03) | 8.23e−03/d | 8.08e−03/d | 5.29e−03/d | 1.17e−02/d | 1.4 | 26 |
| Fails to start (standby) | Gamma(1.72e+01,4.79e−04) | 8.24e−03/d | 8.08e−03/d | 5.27e−03/d | 1.18e−02/d | 1.5 | 26 |

with missing data, a Bayesian result has been proposed. By assuming a constant failure rate and an exponential failure-time distribution, this approach models the uncertainty associated with the failure-rate parameter using a Gamma distribution. Considering the same parametric family of distributions (Gamma in our case) for the failure rate of all components and failure modes, regardless of the database used, makes the theoretical development for subsequent fusion more flexible. The results obtained from fitting Gamma distributions are highly consistent with the available data.

In the case where only the mean value of the failure rate is known, the error factor is fixed to 5 in accordance with [14] here. However, it would be more appropriate, for these cases, to consider an exponential distribution with a parameter whose mean matches the value reported in the databases, so that the error factor would always be equal to 4.33 approximately. It is also observed in the analyzed data that the error factors are fairly homogeneous for each failure mode of a given component. Therefore, in cases where the EF is not specified, it could be more consistent to assign a value similar to those observed for that component and failure mode (e.g., the mean value).

Secondly, in relation to the fusion of Gamma failure rate distributions coming from multiple data sources for the same component, two different methodologies have been proposed. The first approach is based on the method of moments, which consists of considering the Gamma distribution that has the same first and second order moments as the mixture. The second is a data-driven approach that takes the entire distribution into account and estimates the Gamma parameters by minimizing the Kullback–Leibler divergence. The simulation studies reveal that the results are very good in the homogeneous case, even for small sample sizes (number of data sources for each component). In the heterogeneous case, a much larger sample size is required to accurately estimate the dominant population distribution. For this reason, the method of moments represents an appropriate and computationally efficient approach under homogeneous conditions, whereas a methodology based on Kullback–Leibler divergence is preferable when heterogeneity among data sources is present or cannot be established a priori, and therefore, is adopted as the baseline method in such cases. Besides, the analyzed data show a high degree of homogeneity among the data corresponding to the same component and failure mode. Therefore, it would be advisable to first explore the results of the fusion by failure mode and, subsequently, if necessary, to study the fusion of the resulting heterogeneous distributions.

**Table 8**
Estimates for the pressure sensor component. All failure rates are expressed in failures per hour (/h).

| Failure Mode | Probability Distribution Function (PDF) | $\lambda_{mean}$ (Ref.Mean) | $\lambda_{median}$ ($Md$) | $\lambda_{5\%}$ ($LB$) | $\lambda_{95\%}$ ($UB$) | Error Factor ($EF$) | # |
|---|---|---|---|---|---|---|---|
| Fail to operate | Gamma | 8.22e−07/h | 3.74e−07/h | 3.23e−09/h | 3.16e−06/h | 8.4 | 27 |
| Fail to operate | Gamma(5e−01,1.64e−06) | 8.20e−07/h | 3.73e−07/h | 3.22e−09/h | 3.15e−06/h | 8.4 | 27 |
| Fail to operate | Gamma(5.00e−01,1.64e−06) | 8.22e−07/h | 3.74e−07/h | 3.22e−09/h | 3.16e−06/h | 8.4 | 27 |
| Failure to funct./oper. | Gamma | 1.80e−07/h | | | | 5 | 28 |
| Failure to funct./oper. | Gamma(8.4e−01,2.14e−07) | 1.80e−07/h | 1.16e−07/h | 5.73e−09/h | 5.73e−07/h | 5 | 28 |
| Failure to funct./oper. | Gamma(8.33e−01,2.16e−07) | 1.80e−07/h | 1.15e−07/h | 5.58e−09/h | 5.76e−07/h | 5 | 28 |
| Failure to funct./oper. | Gamma | 8.80e−07/h | | | | 5 | 29 |
| Failure to funct./oper. | Gamma(8.4e−01,1.05e−06) | 8.83e−07/h | 5.67e−07/h | 2.81e−08/h | 2.81e−06/h | 5 | 29 |
| Failure to funct./oper. | Gamma(8.33e−01,1.06e−06) | 8.80e−07/h | 5.63e−07/h | 2.73e−08/h | 2.81e−06/h | 5 | 29 |
| Failure to funct./oper. | Gamma | 1.80e−06/h | | | | 5 | 30 |
| Failure to funct./oper. | Gamma(1.02e+00,3.3e−06) | 3.36e−06/h | 2.34e−06/h | 1.80e−07/h | 1.00e−05/h | 4.3 | 30 |
| Failure to funct./oper. | Gamma(8.33e−01,2.16e−06) | 1.80e−06/h | 1.15e−06/h | 5.58e−08/h | 5.76e−06/h | 5 | 30 |
| Failure to funct./oper. | Gamma | 2.50e−06/h | | | | 5 | 31 |
| Failure to funct./oper. | Gamma(8.4e−01,2.97e−06) | 2.50e−06/h | 1.60e−06/h | 7.95e−08/h | 7.96e−06/h | 5 | 31 |
| Failure to funct./oper. | Gamma(8.33e−01,3.00e−06) | 2.50e−06/h | 1.60e−06/h | 7.75e−08/h | 7.99e−06/h | 5 | 31 |
| Failure to funct./oper. | Gamma | 6.10e−07/h | | | | 5 | 32 |
| Failure to funct./oper. | Gamma(1.02e+00,1.12e−06) | 1.14e−06/h | 7.94e−07/h | 6.08e−08/h | 3.39e−06/h | 4.3 | 32 |
| Failure to funct./oper. | Gamma(8.33e−01,7.32e−07) | 6.10e−07/h | 3.90e−07/h | 1.89e−08/h | 1.95e−06/h | 5 | 32 |
| Failure to funct./oper. | Lognormal | 1.00e−05/h | | | | 10 | 33 |
| Failure to funct./oper. | Lognormal(−1.25e+01,1.4e+00) | 1.00e−05/h | 3.75e−06/h | 3.75e−07/h | 3.75e−05/h | 10 | 33 |
| Failure to funct./oper. | Gamma(4.40e−01,2.28e−05) | 1.00e−05/h | 4.02e−06/h | 1.89e−08/h | 4.02e−05/h | 10 | 33 |
| Fails to funct. on demand | Unspecified | 7.10e−07/h | | | | 5 | 34 |
| Fails to funct. on demand | Lognormal(−1.51e+01,1.4e+00) | 7.10e−07/h | 2.67e−07/h | 2.67e−08/h | 2.67e−06/h | 10 | 34 |
| Fails to funct. on demand | Gamma(8.33e−01,8.52e−07) | 7.10e−07/h | 4.54e−07/h | 2.20e−08/h | 2.27e−06/h | 5 | 34 |
| Failure to funct./oper. | Unspecified | 4.60e−07/h | | | | 5 | 35 |
| Failure to funct./oper. | Lognormal(−1.56e+01,1.4e+00) | 4.60e−07/h | 1.73e−07/h | 1.73e−08/h | 1.73e−06/h | 10 | 35 |
| Failure to funct./oper. | Gamma(8.33e−01,5.52e−07) | 4.60e−07/h | 2.94e−07/h | 1.43e−08/h | 1.47e−06/h | 5 | 35 |

**Table 9**
Integrating the probability distributions estimated for each component displayed in Table 1 in [14] by means of moments and KL methodologies.

| Component | Moments | KL divergence |
|---|---|---|
| Interface flange | $\Gamma(\hat{\alpha} = 0.157, \hat{\beta} = 8.419 \times 10^{-6})$ | $\Gamma(\hat{\alpha} = 0.316, \hat{\beta} = 4.173 \times 10^{-6})$ |
| Vacuum pump | $\Gamma(\hat{\alpha} = 0.394, \hat{\beta} = 4.546 \times 10^{-4})$ | $\Gamma(\hat{\alpha} = 0.714, \hat{\beta} = 2.505 \times 10^{-4})$ |
| RF amplifier | $\Gamma(\hat{\alpha} = 0.144, \hat{\beta} = 0.090)$ | $\Gamma(\hat{\alpha} = 0.172, \hat{\beta} = 0.075)$ |
| HV transformer | $\Gamma(\hat{\alpha} = 0.509, \hat{\beta} = 4.470 \times 10^{-4})$ | $\Gamma(\hat{\alpha} = 0.400, \hat{\beta} = 5.694 \times 10^{-4})$ |
| Helium compressor | $\Gamma(\hat{\alpha} = 0.663, \hat{\beta} = 0.028)$ | $\Gamma(\hat{\alpha} = 0.9417, \hat{\beta} = 0.020)$ |
| Pressure sensor | $\Gamma(\hat{\alpha} = 0.113, \hat{\beta} = 1.775 \times 10^{-5})$ | $\Gamma(\hat{\alpha} = 0.425, \hat{\beta} = 4.699 \times 10^{-5})$ |

It should also be emphasized that, within the proposed framework, the management of heterogeneity is primarily addressed before the statistical fusion stage. As described in Section 2.1, the selection of the engineering records to be included in the analysis is carried out by a panel of experts who assess the technical, operational, and exploitation similarity between the candidate information sources and the target application. The purpose of this preliminary engineering assessment is to exclude data sources that are not sufficiently comparable, thereby ensuring that only relevant records are considered for statistical fusion. Nevertheless, some degree of residual heterogeneity may still remain owing to unavoidable differences in operating conditions, data quality, or incomplete knowledge. In such situations, comparing the fitted Gamma distributions prior to fusion (for example, through graphical inspection or by means of distributional distance measures) may help identify information sources requiring further engineering review before they are incorporated into the final unified distribution. In any case, whenever homogeneity cannot be verified a priori, the KL method is recommended as the baseline.

From a maintenance planning perspective, the unified Gamma distributions reported in Table 9 are intended to serve as direct inputs to optimize the preventive maintenance programme of IFMIF-DONES. In this regard, [13] have compiled a Maintenance Matrix comprising 1246 preventive maintenance operations, whose optimal scheduling, resource allocation, and spare parts provisioning strongly rely on component-level failure rate estimates. By providing a single, theoretically grounded probability distribution per component — derived from the systematic fusion of all available data sources — the proposed methodology enables bridging the gap between heterogeneous reliability databases and the operational maintenance planning, thereby contributing to a consistent and traceable RAMI analysis across all subsystems of the facility.

Furthermore, the integrated failure rate distributions have direct implications for spare parts management. As noted in [13], the combination of failure time estimates with procurement lead times is essential for anticipating the purchase of spare parts and for managing the obsolescence of components, particularly those with limited or no prior operational history such as prototype equipment. In this context, the unified Gamma distributions produced by the proposed methodology can serve as inputs to spare parts provisioning models, enabling the estimation of the expected number of failures over a given planning horizon and, consequently, the determination of optimal stock levels and reorder points.

Another important problem is the weight selection in probabilistic fusion. The most straightforward approach assigns equal weights to all densities as in our case. Different strategies could be proposed to specify the weights in the mixture. They may be elicited from expert knowledge, determined through optimization-based approaches, or inferred within a Bayesian framework. In this sense, a hierarchical Bayesian approach that takes into account extra knowledge by introducing prior distributions on the model parameters might be a suitable option. Here, it would be very important to assess its behavior as a function of the available sample size and its heterogeneity. In addition, the formulation of alternative models with different priors and hyperpriors, selected on the basis of available expert knowledge, would also be required.

Finally, complete implementation using open-source software guarantees full computational traceability, facilitating cross-validation and future methodological extensions without dependence on proprietary software.

In conclusion, the theoretical framework presented in this study is clear and concise and enables the derivation of coherent initial guidelines to support a unified design of DONES Accelerator Systems. In particular, its application to a substantially larger dataset, such as those available from the RAMI analyses within the frame of the EUROfusion Early Neutron Source Work Package (WPENS), would be a valuable next step in the sector.

## 6. Conclusions

This work presents a novel methodology to combine heterogeneous reliability datasets associated with fusion-specific devices. The algorithm proposed in this work aims to support the RAMI studies for fusion-oriented research installations, for which the highest requirements on the system availability are defined. This, in turn, is part of the reliability-centered engineering design activities of the emerging fusion systems. In this sense, the method developed herein and used for equipment of the IFMIF-DONES accelerator systems applies to the problem of design optimization for DONES. It also addresses the more general challenge of handling multi-source datasets on the reliability of fusion-specific components in the design process of novel fusion facilities under development.

The procedure consists of three main phases. First, experts systematically select relevant data records from multiple sources using a structured and uniform approach. Then, the failure-rate distribution is estimated from incomplete data using the well-known Poisson-Gamma framework, resulting in an estimated Gamma distribution. Two methods are then proposed in the third phase to fuse these Gamma distributions: the method of moments and a method based on minimizing the Kullback–Leibler divergence. Both methods provide promising results by estimating the parameters of the resulting Gamma distribution from the original data, although the simulation results indicate that the method of moments is efficient under homogeneous conditions, whereas the Kullback–Leibler approach performs better with heterogeneous data. The methodology is fully implemented using open-source software, ensuring computational traceability. In general, the proposed framework provides clear guidelines for the unified design of DONES accelerator systems and can be applied for a wide range of fusion-oriented research facilities.

## CRediT authorship contribution statement

**Helena Ortiz:** Writing – review & editing, Writing – original draft, Visualization, Validation, Software, Methodology, Formal analysis, Data curation. **Juan Eloy Ruiz-Castro:** Writing – review & editing, Writing – original draft, Visualization, Validation, Supervision, Project administration, Methodology, Investigation, Funding acquisition, Formal analysis, Data curation. **Christian Acal:** Writing – review & editing, Writing – original draft, Visualization, Validation, Supervision, Software, Methodology, Investigation, Formal analysis, Data curation. **Karol Kowal:** Writing – review & editing, Writing – original draft, Supervision, Project administration, Investigation, Funding acquisition, Data curation, Conceptualization. **Juan Chiachío:** Writing – review & editing, Writing – original draft, Supervision, Project administration, Investigation, Funding acquisition, Data curation, Conceptualization. **Ana María Aguilera:** Writing – review & editing, Writing – original draft, Visualization, Validation, Supervision, Project administration, Methodology, Investigation, Funding acquisition, Formal analysis, Data curation.

## Declaration of generative AI and AI-assisted technologies in the manuscript preparation process

AI has been used to review writing in the English language.

## Declaration of competing interest

The authors declare that they have no known competing financial interests or personal relationships that could have appeared to influence the work reported in this paper.

## Acknowledgments

The authors acknowledge financial support by the projects PID2023-149087NB-I00 funded by MICIU/AEI/10.13039/501100011033 and by FEDER, UE. This research has also been supported by IMAG María de Maeztu grant CEX2020-001105-M/AEI/ 10.13039/501100011033. In addition, this work has been carried out within the framework of the EUROfusion Consortium, funded by the European Union via the Euratom Research and Training Programme (Grant Agreement No 101052200 - EUROfusion). Views and opinions expressed are, however, those of the author(s) only and do not necessarily reflect those of the European Union or the European Commission. Neither the European Union nor the European Commission can be held responsible for them. Besides, this work was part of the project co-financed by the Polish Ministry of Science and Higher Education under the program entitled International Projects Co-financed. Finally, the data analysis presented in this work was carried out by using the Fusion Component Failure Rate Database (FCFRDB) developed by ENEA for F4E and ITER organizations.

## Data availability

In the interest of transparency and knowledge transfer, the data used in the application and the code scripts implemented in the R statistical software are publicly available in Digibug (the University of Granada Digital Repository) for the research community at the following link: https://hdl.handle.net/10481/114938.

In particular, three R scripts containing the complete implementation of the proposed methods are provided, together with illustrative examples:

1. codeR_parameters of gamma distribution.R: This script contains the methodology described in Section 2.2. An illustrative example is also included using the dataset considered by [14] (see Table 1 in [14], or the first rows highlighted in white in Tables 3–8 of the current manuscript), the parameter estimates obtained for the proposed Gamma distribution are reproduced (third rows highlighted in blue in Tables 3–8 of the current manuscript).

2. codeR_gamma_moments_method.R: This script contains the approach described in Section 2.3.1. An illustrative example is provided to show how, using the Gamma distribution parameter estimates obtained in Section 2.2 for the interface component, the result reported in Table 9 of the current manuscript for this component can be reproduced, i.e., $\Gamma(\hat{\alpha} = 0.157, \hat{\beta} = 8.419 \times 10^{-6})$.

3. codeR_Kullback–Leibler divergence.R: This script contains the approach described in Section 2.3.2. Similarly, an illustrative example is provided to show how, using the Gamma distribution parameter estimates obtained in Section 2.2 for the interface component, the result reported in Table 9 of the current manuscript for this component can be reproduced, i.e., $\Gamma(\hat{\alpha} = 0.316, \hat{\beta} = 4.173 \times 10^{-6})$..